\documentclass[preprint2]{aastex}    

\shorttitle{ULX in NGC\,4088}
\shortauthors{Mezcua et al.}

\begin{document}

\title{Revealing the nature of the ULX and X-ray population \\of the spiral galaxy NGC\,4088}

\author{M.~Mezcua\altaffilmark{1,2}, G.~Fabbiano\altaffilmark{3}, J.C.~Gladstone\altaffilmark{4}, S.A.~Farrell\altaffilmark{5}, and R.~Soria\altaffilmark{6}}

\affil{$^{1}$Instituto de Astrof\'isica de Canarias (IAC), E-38200 La Laguna, Tenerife, Spain;  mmezcua@iac.es}
\affil{$^{2}$Universidad de La Laguna, Dept. Astrof\'isica, E-38206 La Laguna, Tenerife, Spain} 
\affil{$^{3}$Harvard-Smithsonian Center for Astrophysics (CfA); 60 Garden Street Cambridge, Massachusetts 02138 USA}
\affil{$^{4}$Department of Physics, University of Alberta, 11322 - 89 Avenue, Edmonton, Alberta, T6G 2G7, Canada}
\affil{$^{5}$Sydney Institute for Astronomy (SIfA), School of Physics, The University of Sydney, NSW 2006, Australia}
\affil{$^{6}$International Centre for Radio Astronomy Research, Curtin University, GPO Box U1987, Perth, WA 6845, Australia}


\begin{abstract}
We present the first \textit{Chandra} and \textit{Swift} X-ray study of the spiral galaxy NGC\,4088 and its ultraluminous X-ray source (ULX N4088--X1). We also report very long baseline interferometry (VLBI) observations at 1.6 and 5 GHz performed quasi-simultaneously with the \textit{Swift} and \textit{Chandra} observations, respectively. 
Fifteen X-ray sources are detected by \textit{Chandra} within the D25 ellipse of NGC\,4088, from which we derive the X-ray luminosity function (XLF) of this galaxy. We find the XLF is very similar to those of star-forming galaxies and estimate a star-formation rate of 4.5 $M_{\odot}$ yr$^{-1}$. 
The \textit{Chandra} detection of the ULX yields its most accurate X-ray position, which is spatially coincident with compact radio emission at 1.6 GHz. The ULX \textit{Chandra} X-ray luminosity, $L_\mathrm{0.2-10.0 keV} = 3.4\ \times$ 10$^{39}$ erg s$^{-1}$, indicates that N4088--X1 could be located at the high-luminosity end of the high-mass X-ray binary (HMXB) population of NGC\,4088. The estimates of the black hole mass and ratio of radio to X-ray luminosity of N4088--X1 rule out a supermassive black hole nature. 
The \textit{Swift} X-ray spectrum of N4088--X1 is best described by a thermal Comptonization model and presents a statistically significant high-energy cut-off. We conclude that N4088--X1 is most likely a stellar remnant black hole in a HMXB, probably fed by Roche lobe overflow, residing in a super-Eddington ultraluminous state. The 1.6 GHz VLBI source is consistent with radio emission from possible ballistic jet ejections in this state. 
\end{abstract}


\section{Introduction}
Early X-ray observations of nearby galaxies with the \textit{Einstein Observatory}, and later with \textit{ROSAT}, \textit{ASCA}, \textit{XMM-Newton}, and \textit{Chandra}, revealed the presence of a population of off-nuclear X-ray point sources with X-ray luminosities above the Eddington limit for a stellar-mass black hole ($L_\mathrm{X} > 10^{39}$ erg s$^{-1}$; \citealt{1989ARA&A..27...87F}). These high luminosities imply black holes (BHs) of masses $>$ 10 $M_{\odot}$, if the accretion is sub-Eddington and the radiation is isotropic. This suggested that ULXs could be intermediate mass BHs (IMBHs) with BH masses in the range $100 \leq M_\mathrm{BH} \leq 10^{5} M_{\odot}$ (e.g., \citealt{1999ApJ...519...89C}). The alternative is stellar-remnant BHs ($M_\mathrm{BH} < 100 M_{\odot}$) accreting at around or above the Eddington limit (e.g., \citealt{2001ApJ...552L.109K}; \citealt{2005Sci...307..533F}; \citealt{2011NewAR..55..166F}). 

Studies of the X-ray luminosity function (XLF) of the X-ray populations of these nearby galaxies are a very useful tool for discerning between different types of X-ray populations and galaxy properties (e.g., morphological type, age of stellar population) and investigating the nature of the X-ray sources that populate them (e.g., X-ray binaries -XRBs; see review by \citealt{2006ARA&A..44..323F}). The XLF of late-type star-forming galaxies is usually described by a straight power-law and is associated with a population of high-mass X-ray binaries (HMXBs) located in regions of star formation (e.g., \citealt{2002ApJ...577..726Z}; \citealt{2004ApJ...602..231C}; \citealt{2007ApJ...661..135Z}). Early-type galaxies are best described by an XLF with a broken power-law and a break at a few times 10$^{38}$ erg s$^{-1}$ (attributed to the Eddington luminosity of neutron star XRBs; e.g., \citealt{2000ApJ...544L.101S,2001ApJ...556..533S}), and are thought to be dominated by a population of low-mass XRBs (LMXBs, e.g., \citealt{2004ApJ...611..846K,2010ApJ...721.1523K}; \citealt{2004MNRAS.349..146G}). Studies to the XLF of different galaxies can also be used to gain insight into the nature of ULXs. 

ULX population studies have revealed that many ULXs are associated with young star-forming regions and star-forming galaxies (e.g., \citealt{2004ApJS..154..519S,2009ApJ...703..159S}). They also tend to be found in low-metallicity regions (e.g., \citealt{2009MNRAS.400..677Z}; \citealt{2010ApJ...714.1217B}; \citealt{2010MNRAS.408..234M}). Such low metallicities are required in order to make more massive BHs (e.g., \citealt{2003ApJ...591..288H}), which shows that it would be possible for ULXs to be massive stellar remnant BHs. 
Yet, population studies of the HMXB population of the local universe has also revealed an unbroken power-law slope in the XLF up to $\sim2 \times 10^{40}$ erg s$^{-1}$ (e.g., \citealt{2003MNRAS.339..793G}; \citealt{2011ApJ...741...49S}; \citealt{2012MNRAS.419.2095M}). This continuation indicates that the majority of ULXs with $L_\mathrm{X} < 2 \times 10^{40}$ erg s$^{-1}$ are likely to be HMXBs with a stellar-mass BH accreting at around or above the Eddington limit.

Many studies have been undertaken to obtain the mass of the putative BH contained in ULXs. Such methods include: studying the ULX optical counterpart (e.g., \citealt{2006IAUS..230..293P}; \citealt{2009ApJ...697..950K}; \citealt{2011ApJ...728L...5C}; \citealt{2011ApJ...737...81T}; \citealt{2013ApJS..206...14G}); from X-ray analysis, using either spectral fitting (e.g., \citealt{2003ApJ...585L..37M}; \citealt{2009MNRAS.397..124G}; \citealt{2010MNRAS.402.2559C}; \citealt{2011ApJ...734..111D}; \citealt{2012ApJ...752...34G}; \citealt{2013arXiv1310.0745B}), the luminosity-temperature relation (e.g., \citealt{2003ApJ...585L..37M}; \citealt{2009MNRAS.397..124G}; \citealt{2009ApJ...692..443S}; \citealt{2011ApJ...743....6S}), quasi-periodic oscillations (e.g., \citealt{2003ApJ...586L..61S}; \citealt{2007ApJ...660..580S}; \citealt{2013MNRAS.tmp.2483C}), or X-ray variability (\citealt{2004A&A...423..955S}; \citealt{2009MNRAS.397.1061H}; \citealt{2013MNRAS.tmp.2522D}). 
In those cases where candidate radio counterparts have been identified, attempts have been made to obtain the BH mass using the Fundamental Plane of accreting BHs (e.g., \citealt{2006A&A...452..739S}; \citealt{2011IAUS..275..325C}; \citealt{2011AN....332..379M}; \citealt{2012Sci...337..554W}; \citealt{2013MNRAS.436.1546M}a; \citealt{2013MNRAS.436.3128M}c). The Fundamental Plane is a correlation between 2--10 keV X-ray luminosity, 5 GHz radio luminosity, and BH mass that holds for sub-Eddington accreting BHs in the low/hard X-ray state and with steady jet emission (e.g., \citealt{2003MNRAS.345.1057M}; \citealt{2006A&A...456..439K}; \citealt{2012MNRAS.423..590G}). The detection of compact core radio emission is required in order to locate a ULX in the Fundamental Plane, which can be achieved only by means of very long baseline interferometry (VLBI) radio observations. \cite{2011AN....332..379M} initiated a program with the European VLBI Network (EVN\footnote{www.evlbi.org}) aimed at detecting and studying the milliarcsecond-scale emission of ULXs. Such investigations have indicated that those ULXs with radio counterparts may be powered by IMBHs (e.g., \citealt{2011AN....332..379M}; \citealt{2013MNRAS.436.1546M}a, \citeyear{2013MNRAS.436.3128M}c). Here, we investigate one of these ULXs in NGC\,4088.
		
NGC\,4088 is an asymmetric spiral galaxy ($D_\mathrm{L}$ = 13 Mpc, redshift = 0.002524; \citealt{2001A&A...370..765V}) hosting a ULX (N4088--X1) offset $\sim$32 arcsec from the nucleus. N4088--X1 was first detected with the \textit{ROSAT} satellite by \cite{2005ApJS..157...59L}, who reported an X-ray luminosity $L_\mathrm{0.3-8.0 keV} \sim 6 \times 10^{39}$ erg s$^{-1}$.
The ULX is located within the spiral arm of NGC\,4088, possibly within an HII region (e.g., \citealt{2005ApJS..157...59L}; \citealt{2006A&A...452..739S}). A cross-match of the VLA\footnote{Very Large Array of the National Radio Astronomy Observatory (NRAO).} FIRST\footnote{Faint Images of the Radio Sky at Twenty-cm survey.} catalog with \textit{ROSAT} ULX catalogs (e.g., \citealt{2005ApJS..157...59L}) revealed a 1.4 GHz VLA radio counterpart for the ULX of 1.87 mJy, with an offset between the \textit{ROSAT} and the radio position of 3.6 arcsec (\citealt{2006A&A...452..739S}). Later EVN observations at 1.6 GHz yielded the detection of a compact, unresolved component of 0.1 mJy centered at RA(J2000) = 12$^h$05$^m$31$^s$.7110 $\pm$ 0$^s$.0003, Dec.(J2000) = 50$^{\circ}$32\arcmin46\arcsec.729 $\pm$ 0\arcsec.002 (\citealt{2011AN....332..379M}). This provided an upper limit on the 5 GHz luminosity (assuming a radio spectral index $\alpha$ = 0.15), from which an upper limit on the ULX BH mass of $\sim10^{5} M_{\odot}$ was estimated using the Fundamental Plane of accretion (\citealt{2011AN....332..379M}). 

In this paper we present the first \textit{Chandra} observations of the galaxy NGC\,4088 and the ULX N4088--X1, as well as quasi-simultaneous VLBI observations with the EVN at 5 GHz. We also report a reanalysis of the EVN data at 1.6 GHz, now imaged at the \textit{Chandra} position reported in this paper, and the analysis of 23 \textit{Swift} observations performed nearly at the same time (between April and September 2009) as the 1.6 GHz EVN observations (2009 June 1--2). With these data we investigate the proposed association between the X-ray and radio emission (for which a sub-arcsecond X-ray position is needed) of N4088--X1 and attempt to estimate the BH mass with the aim of revealing the nature of this ULX. 

The paper is organized as follows: the observations and data analysis are presented in Section~\ref{observations}, while the main results obtained are shown in Section~\ref{results} and discussed in Section~\ref{discussion}. Final conclusions are presented in Sect.~\ref{conclusions}. 

Through this paper we assume a $\Lambda$ CDM cosmology with parameters $H_\mathrm{0} = 73$ km s$^{-1}$ Mpc$^{-1}$, $\Omega_{\Lambda}=0.73$, and $\Omega_\mathrm{m}=0.27$.

\section{Observations and data analysis}
\label{observations}
\subsection{\textit{Chandra} X-ray observations}
\label{chandraobservations}
The ULX N4088--X1 was observed on 2012 June 06 (Obs. ID: 14442; PI: Mezcua) with the \textit{Chandra X-ray observatory} (Weisskopf et al. 2002). The observation was performed using the Advanced CCD Imaging Spectrometer detector (\citealt{1997AAS...190.3404G}; ACIS-S) with an integration time of 19.8 ks. The data were reprocessed using CIAO version 4.5 and the corresponding calibration files, following the standard \textit{Chandra} ACIS data analysis threads\footnote{http://cxc.harvard.edu/ciao/threads/}. The \textit{chandra\_repro} reprocessing script was used to reprocess the data and generate a new level=2 event file. 

An image of the S3 chip (ccd\_id=7), a background image, and a PSF map that provides the size of the PSF at each pixel in the image were then produced using the tools \textit{dmcopy}, \textit{aconvolve}, and \textit{mkpsfmap} and given to \textit{wavdetect}, which performed source detection and extracted source net counts in the energy range 0.3--10 keV. 
The detected source count rates for all sources lying within the D25 ellipse of NGC 4088 were converted to source fluxes by applying a conversion factor calculated assuming a power-law spectrum of $\Gamma$ = 1.8 and line-of-sight Galactic absorption\footnote{$N_\mathrm{H}$ calculated using the COLDEN tool: http://cxc.harvard.edu/toolkit/colden.jsp} $N_\mathrm{H} = 2 \times 10^{20}$ cm$^{-2}$ using WebPIMMS\footnote{http://heasarc.nasa.gov/Tools/w3pimms.html}. A $\Gamma$ = 1.4$\sim$2 (e.g., \citealt{2002ApJ...577..726Z}) and $\Gamma$ = 1.4--1.8 (e.g., \citealt{2011ApJ...729...12B}) was found for HMXBs and LMXBs, respectively. We note that the average value of $\Gamma$ = 1.8 used here may not be applicable to our case
since those studies have a deeper detection limit that includes also XRBs in the low/hard state (dominated by a hard power-law component), while we can only see those in the high/soft state (dominated by a disk blackbody). 
To make sure that we are using a plausible slope, we used the {\footnotesize XSPEC} task {\it fakeit} to simulate a distribution of {\it Chandra} spectra for disk-blackbody models with column densities equal to the line-of-sight and twice the line-of-sight values, and inner-disk temperatures distributed between 0.6 and 1.0 keV (typical disk temperatures in the high/soft state; e.g., \citealt{2006csxs.book..157M}).  We then re-fitted the simulated spectra with a power-law model, fixing the column density to the line-of-sight value and leaving the photon index as a free parameter. We find that low signal-to-noise disk-blackbody spectra in that temperature range are indeed approximated in the {\it Chandra} band (at least for the purpose of converting from count rates to fluxes) by power-laws of photon index between $\sim$1.5 (for $T_\mathrm{in} \sim$ 1 keV) and $\sim$2 (for $T_\mathrm{in} \sim$ 0.6 keV), with the most likely value $\Gamma$ $\sim$ 1.8.

The source spectrum of N4088--X1 was extracted using the \textit{specextract} script and selecting a circular region of 2$\arcsec$ around the target source and of 15$\arcsec$ in a source-free area of the same chip for the background. The extracted spectrum was grouped to 15 counts per energy bin to allow for $\chi^{2}$ fitting using the tool \textit{grppha}.

\subsection{\emph{Swift} X-ray Observations}
NGC\,4088 was observed on 23 occasions in X-rays with the \emph{Swift} X-ray Telescope (XRT) between 2009 April 15 and 2009 September 21 as part of a monitoring campaign targeting the supernova (SN) SN2009dd (see Table \ref{swiftlog} for a log of the \emph{Swift} observations). We generated images, light curves and spectra, including the background and ancillary response files, with the online XRT data product generator \footnote{http://swift.ac.uk/user\_objects/} (\citealt{2009MNRAS.397.1177E}). We downloaded suitable spectral response files for single and 
double events in photon-counting mode from the latest calibration database. N4088--X1 was clearly detected in all observations with an average XRT count rate of 0.0037 count s$^{-1}$. We extracted a light curve binned at the duration of each individual observation, finding evidence for small deviations from the average count rate in 5 of the 23 observations, though no evidence for significant spectral variability could be seen in the standard \emph{Swift hardness ratio ((1.5--10
keV)/(0.3--1.5 keV))}. We next extracted a combined spectrum from the \emph{Chandra} position of N4088--X1, excluding the 5 observations in which evidence for variability was seen, giving a total exposure time of 76 ks. 

\begin{table}[h!]
\caption{Log of the \emph{Swift} XRT observations of N4088--X1. The observations used for producing the combined spectrum are indicated in column 4.\label{swiftlog}} 
\scriptsize
\begin{tabular}{cccc}
\tableline
Date  & Obsid & Exp. time (s)  & Spectrum\\
\tableline
2009-04-15 & 00031401001 & 1996 & Y \\
2009-04-17 & 00031401002 & 830 & Y \\
2009-04-19 & 00031401003 & 3946 & Y \\
2009-04-21 & 00031401004 & 2670 & Y \\
2009-04-25 & 00031401005 & 2712 & Y \\
2009-05-10 & 00031401006 & 4306 & Y \\
2009-05-16 & 00031401007 & 6045 & Y \\
2009-07-04 & 00031401008 & 3824 & Y \\
2009-07-05 & 00031401009 & 5957 & Y \\
2009-07-15 & 00031401011 & 3930 & N \\
2009-07-16 & 00031401012 & 8963 & N \\
2009-07-19 & 00031401013 & 4863 & Y \\
2009-07-19 & 00031401014 & 6049 & N \\
2009-07-26 & 00031401015 & 4668 & N \\
2009-07-27 & 00031401016 & 5181 & N \\
2009-08-02 & 00031401017 & 5934 & Y \\
2009-08-03 & 00031401018 & 4647 & Y \\
2009-08-09 & 00031401019 & 6968 & Y \\
2009-09-13 & 00031401020 & 33 & Y \\
2009-09-14 & 00031401021 & 4593 & Y \\
2009-09-16 & 00031401022 & 4587 & Y \\
2009-09-20 & 00031401023 & 5858 & Y \\
2009-09-21 & 00031401024 & 6135 & Y \\
\tableline
\end{tabular}
\end{table}

\subsection{VLBI radio observations}
\label{VLBI}
N4088--X1 was observed with the EVN at 5 GHz on 2012 June 01 (project code: EM095A; PI: Mezcua).
Eight antennas participated in the observations: Effelsberg (Germany), Westerbork (The Netherlands), Jodrell Bank (United Kingdom), Onsala (Sweden), Medicina (Italy), Noto (Italy), Torun  (Poland), and Yebes (Spain). The observations were performed using the phase-referencing technique, in which the target and a nearby, compact, bright source (the phase calibrator) are observed interleaving scans. A target-phase calibrator cycle of four minutes (3 minutes on the target, 1 min on the phase calibrator) was used. As a result, a total of 2.7 h was spent on the target source. The source J1203+4803 was used as phase calibrator, while the bright and compact radio source 4C+39.25 was used as fringe finder and bandpass calibrator.

The data were recorded in dual-circular polarization and at sample rate of 1024 megabit per second (Mbps). Eight intermediate frequency bands, each of 16 MHz each and 32 spectral channels, were used. The data correlation was performed at JIVE\footnote{Joint Institute for VLBI in Europe, Dwingeloo, the Netherlands.} with an averaging time of 4 sec.

The calibration of the correlated data was performed using AIPS\footnote{Astronomical Image Processing Software of NRAO.}. Amplitudes were calibrated using the gains of the antennas and system temperatures. The data were then fringe-fitted using the phase calibrator. Delay, delay rate, and phase solutions derived from the phase calibrator were interpolated and applied to the target.
The imaging was performed using CLEAN deconvolution with the AIPS task IMAGR. Natural weighting images were produced from non-channel-averaged data\footnote{No channel averaging was applied to the calibrated data to avoid degradation of the synthesized beam away from the phase center (bandwidth smearing).}
and imaging two fields of view (FOV) derived from the positional errors of the VLA and \textit{Chandra} ULX counterparts: one of $\sim$1 $\times$ 1 arcsec$^{2}$ centered at the \textit{Chandra} position, and another one of $\sim$5 $\times$ 5 arcsec$^{2}$ centered at the VLA position. The restoring beam size was 5.7 mas $\times$ 4.5 mas. To enhance the sensitivity, we also repeated the multi-field imaging without using the baselines longer than 20 M$\lambda$. 

We did also recalibrate the 1.6 GHz EVN data previous analyzed by \cite{2011AN....332..379M} and reimaged the data without averaging the channels and using the same two FOVs as at 5 GHz. The imaging was also performed using only the baselines shorter than 15 M$\lambda$, which resulted in a beam size of 33 mas $\times$ 27 mas.

\section{Results}
\label{results}
\begin{figure}[h!]
\includegraphics[scale=0.32]{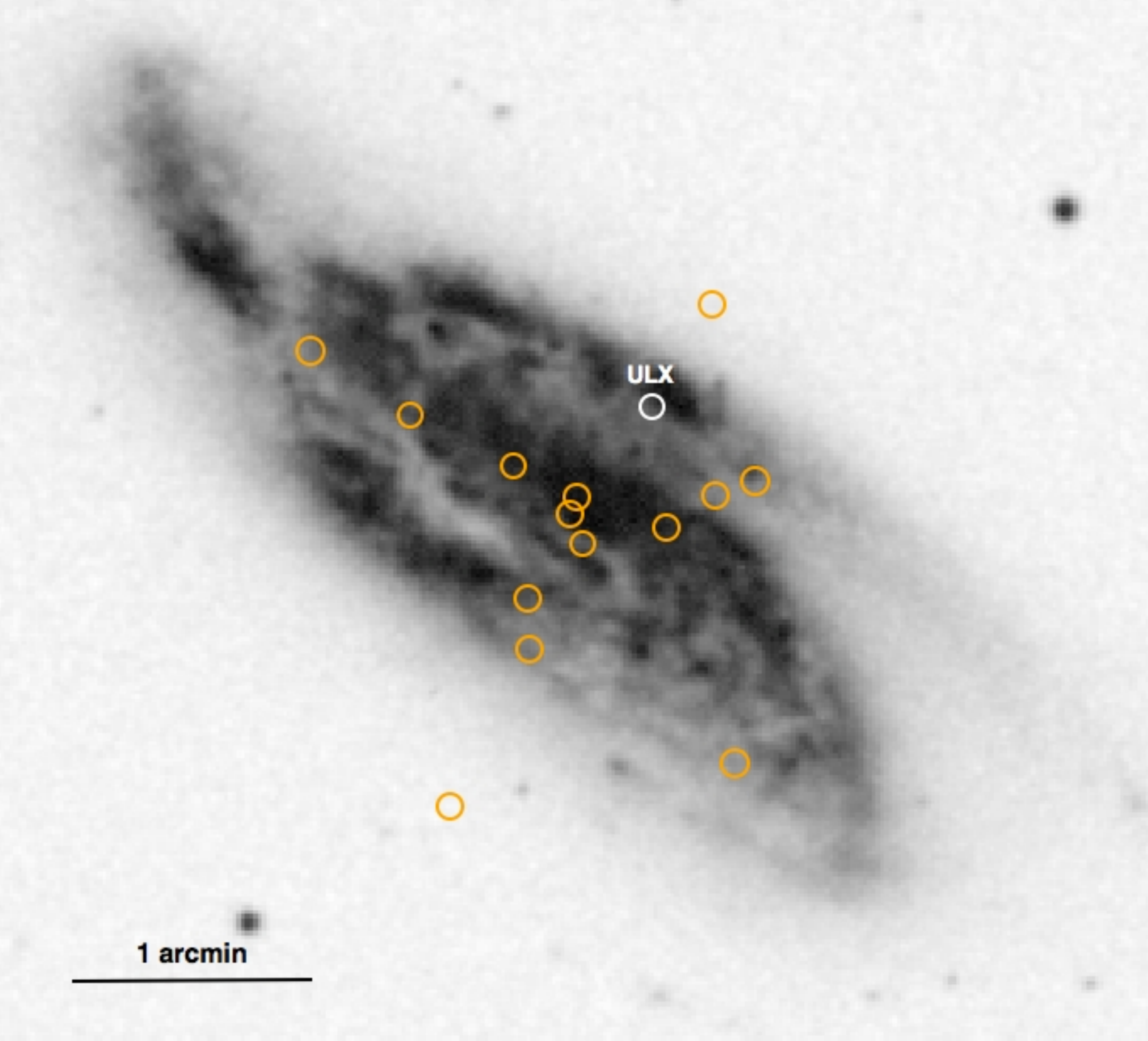}
  \protect\caption{Digitized Sky Survey (\citealt{1990AJ.....99.2019L}) image of NGC\,4088. The position of the fifteen sources detected by \textit{wavdetect} in the 0.3--10 keV band within the D25 ellipse of NGC\,4088 are marked with orange circles of radius 3 arcsec. The position of the ULX is marked with a white circle.\label{fig1}}
\end{figure}

A total of thirty-one X-ray sources are detected by \textit{wavdetect} in the 0.3--10 keV band, of which fifteen lay within the D25 ellipse of the host galaxy. 
The location of the fifteen detected sources is shown in Fig.~\ref{fig1}, and their position, observed flux, and 0.3--10 keV band luminosity are provided in Table~\ref{table1}. 

\begin{table*}[t!]
\begin{center}
\caption{Point-like \textit{Chandra} X-ray sources detected in the 0.3--10 keV band inside the D25 ellipse of NGC\,4088. Fluxes and luminosities are unabsorbed, derived assuming $\Gamma = 1.8$ and line-of-sight Galactic column density $N_\mathrm{H} = 2 \times 10^{20}$ cm$^{-2}$. 
The positional 1$\sigma$-errors reported correspond to the statistical uncertainties from the PSF fitting performed by \textit{wavdetect}. A systematic pointing error of up to 0.6 arcsec can also affect the \textit{Chandra} absolute astrometry. $^{a}$ Src. 4 is the most probable nucleus of NGC\,4088. $^{b}$ Not detected by \textit{wavdetect}, counts obtained from CIAO Statistics. \label{table1}}
\begin{tabular}{lcccc}
\tableline
\tableline
 Source	&  RA			& 	Dec.      		&   Net counts	& $L_\mathrm{0.3-10 keV}$\\
      		&  (J2000)        	&     (J2000)      	&      	       & ($\times 10^{38}$ erg s$^{-1}$ )	\\
\tableline
N4088--X1&	12 05 32.33	$\pm$	0.01	&	50 32 45.9	$\pm$	0.1	&	221.2	$\pm$	15.3	&	21.9	$\pm$	1.5	\\
Src. 2	&	12 05 29.73	$\pm$	0.01	&	50 32 28.3	$\pm$	0.1	&	89.1	$\pm$	9.8	&	8.8	$\pm$	1.0	\\
Src. 3	&	12 05 31.92	$\pm$	0.01	&	50 32 16.9	$\pm$	0.2	&	60.5	$\pm$	8.3	&	6.0	$\pm$	0.8	\\
Src. 4\tablenotemark{a}	&	12 05 34.36	$\pm$	0.02	&	50 32 19.9	$\pm$	0.2	&	49.3	$\pm$	7.8	&	4.9	$\pm$	0.8	\\
Src. 5	&	12 05 35.41	$\pm$	0.01	&	50 31 59.7	$\pm$	0.1	&	49.3	$\pm$	7.4	&	4.9	$\pm$	0.7	\\
Src. 6	&	12 05 30.70	$\pm$	0.02	&	 50 32 24.9	$\pm$	0.2	&	47.3	$\pm$	7.4	&	4.7	$\pm$	0.7	\\
Src. 7	&	12 05 34.18	$\pm$	0.01	&	50 32 24.3	$\pm$	0.1	&	27.6	$\pm$	6.1	&	2.7	$\pm$	0.6	\\
Src. 8	&	12 05 35.34	$\pm$	0.00	&	50 31 47.6	$\pm$	0.1	&	22.8	$\pm$	5.1	&	2.3	$\pm$	0.5	\\
Src. 9	&	12 05 30.83	$\pm$	0.02	&	50 33 10.8	$\pm$	0.2	&	20.3	$\pm$	4.8	&	2.0	$\pm$	0.5	\\
Src. 10	&	12 05 38.40	$\pm$	0.01	&	50 32 43.2	$\pm$	0.1	&	17.1	$\pm$	4.5	&	1.7	$\pm$	0.4	\\
Src. 11	&	12 05 37.29	$\pm$	0.02	&	50 31 09.4	$\pm$	0.4	&	15.1	$\pm$	4.4	&	1.5	$\pm$	0.4	\\
Src. 12	&	12 05 40.92	$\pm$	0.05	&	50 32 58.6	$\pm$	0.3	&	13.9	$\pm$	4.4	&	1.4	$\pm$	0.4	\\
Src. 13	&	12 05 35.80	$\pm$	0.03	&	50 32 31.4	$\pm$	0.3	&	11.6	$\pm$	4.1	&	1.2	$\pm$	0.4	\\
Src. 14	&	12 05 34.02	$\pm$	0.01	&	50 32 12.9	$\pm$	0.2	&	8.2	$\pm$	3.5	&	0.8	$\pm$	0.3	\\
Src. 15	&	12 05 30.13	$\pm$	0.04	&	50 31 21.0	$\pm$	0.2	&	8.0	$\pm$	3.2	&	0.8	$\pm$	0.3	\\
SN2009dd\tablenotemark{b} & 12 05 34.10					&	50 32 19.4				&      8.0   $\pm$	3.2	&	0.8	$\pm$	0.3	\\	
\tableline
\end{tabular}
\end{center}
\end{table*}

\subsection{X-ray properties of the ULX}
\label{xrayprop}
For the ULX N4088--X1, we obtain an X-ray luminosity of $L_\mathrm{0.3-10 keV} = 2.2 \times 10^{39}$ erg s$^{-1}$ using the observed \textit{Chandra} fluxes derived from the count rates assuming the power-law model described in Sect.~\ref{chandraobservations}. The \textit{Chandra} detection of N4088--X1 shows that the source is located at RA(J2000) = 12$^h$05$^m$32$^s$.33 $\pm$ 0$^s$.01, Dec.(J2000) = 50$^{\circ}$32\arcmin45\arcsec.9 $\pm$ 0\arcsec.1.
The positional 1$\sigma$-errors correspond to the statistical uncertainties affecting the \textit{wavdetect} centroid algorithm and the dispersion of photons due to the PSF. In addition, the \textit{Chandra} absolute astrometry can be shifted by up to 0.6 arcsec due to pointing uncertainties.

\subsubsection{\textit{Chandra} ULX spectral modeling}
\begin{figure}[h!]
 \hspace{-13pt}
\includegraphics[angle=-90,scale=0.29]{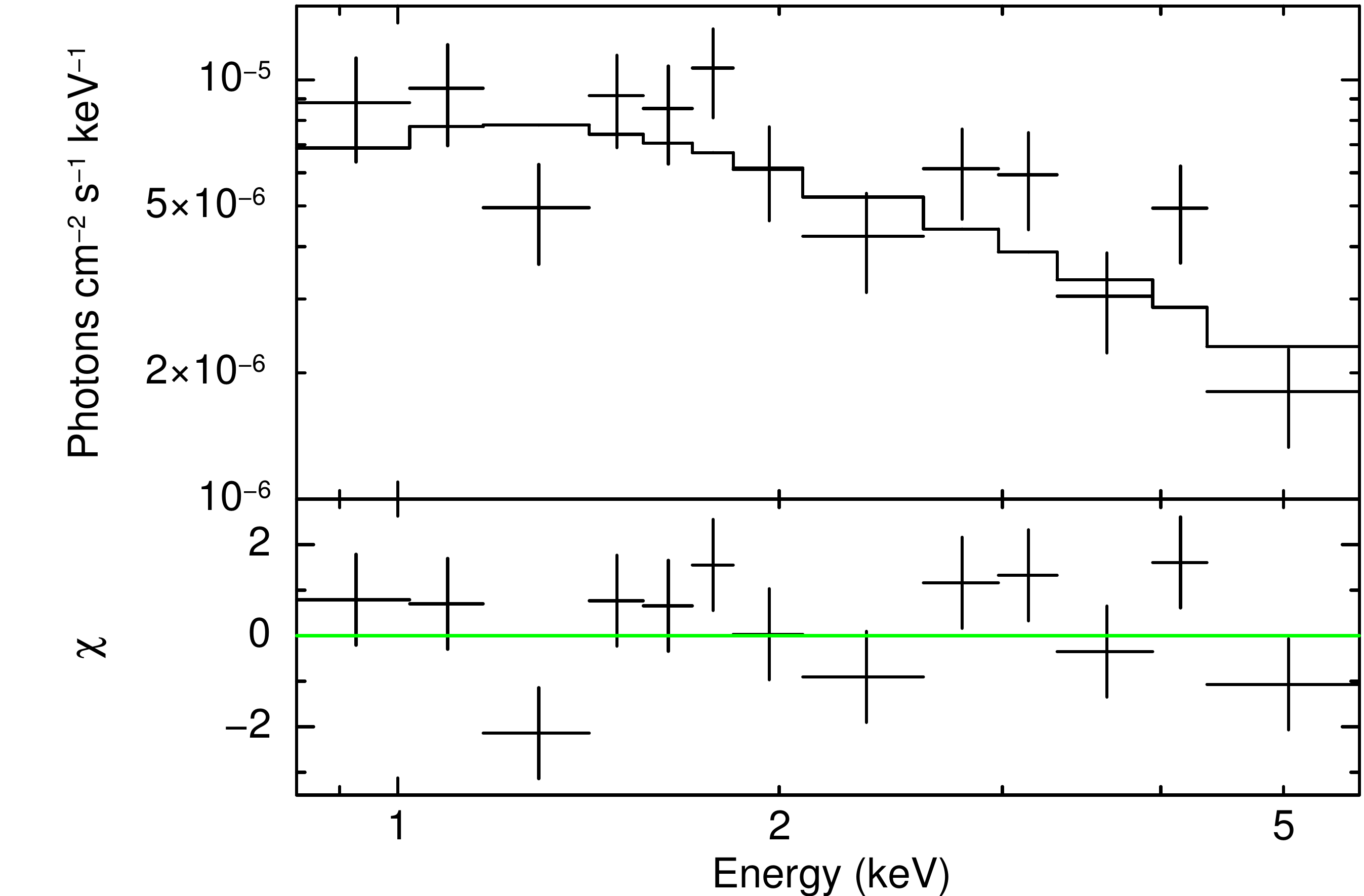}
  \protect\caption{Power-law fit to the \textit{Chandra} spectrum of N4088--X1. \label{chandrapow}}
\end{figure}

We fitted the \textit{Chandra} spectrum using the X-ray spectral fitting package {\footnotesize XSPEC} (\citealt{1996ASPC..101...17A}) v12.7.1 in the 0.2 to 10.0 keV energy range.
Two models were used: an absorbed multicolor disk-blackbody model (\textit{wabs*diskbb}) and a power-law model (\textit{wabs*pow}).
The fits were performed using the minimum $\chi^{2}$ method (i.e. Gehrels Chisq statistics).
The fitting results of each model are shown in Table~\ref{chandra}, while a plot of the power-law spectral fit is shown in Fig.~\ref{chandrapow}.

Both the \textit{pow} and the \textit{diskbb} model provide acceptable statistical fits (i.e. null-hypothesis probability $>$ 5\%, so rejection likelihood $<$ 95\%), with values of $N_\mathrm{H}\sim$ 10--15 times larger than the Galactic column density (i.e. $N_\mathrm{H} = 2 \times 10^{20}$ cm$^{-2}$, \citealt{2003ApJ...588..805K}). The data are insufficient to statistically distinguish between models. 
The disk-blackbody model provides an inner disk temperature $kT_\mathrm{in}$ = 2.5 keV, too high for a standard disk. The power-law fit provides a very hard photon index $\Gamma$ = 1.1 and unabsorbed flux = 1.7 $\times 10^{-13}$ erg s$^{-1}$ cm$^{2}$, from which we derive an X-ray luminosity $L_\mathrm{0.2-10.0 keV}$ = 3.4 $\times\ 10^{39}$ erg s$^{-1}$.  

\begin{table*}
\centering
\caption{Spectral parameters obtained by fitting the \emph{Chandra} spectrum with an absorbed power-law and absorbed disk-blackbody model. The quoted flux is the observed flux in the 0.2 to 10 keV energy range. The errors provided are at 90\% confidence level. \label{chandra}}
  \begin{tabular}[alignment]{lcccccc}
  \hline  
Model  & $N_\mathrm{H}$ & $\Gamma/T_\mathrm{in}$ & Norm. & Flux & $\chi^2$/dof & p-value\\
 & (10$^{22}$ cm$^{-2}$) & ( /keV) &  & (erg cm$^{-2}$ s$^{-1}$) & & \\
\hline
\textit{pow} & 0.3$^{+0.5}_{-0.3}$ & 1.1$^{+0.6}_{-0.5}$ & (1.6$^{+2}_{-0.7}$) $\times$ 10$^{-5}$ & (1.7$^{+0.6}_{-1.5}$) $\times$ 10$^{-13}$ & 16.9/10 & 0.08\\
\textit{diskbb} & 0.2$^{+0.3}_{-0.2}$ & 2.5$^{+4}_{-0.9}$ & 2.1 $\times$ 10$^{-4}$ & (1.4$^{+0.2}_{-1.4}$) $\times$ 10$^{-13}$ & 15.62/10 & 0.11\\
\hline
\end{tabular}
\end{table*}

\subsubsection{\textit{Swift} ULX spectral modeling}
\begin{figure}[h!]
  \includegraphics[scale=0.31]{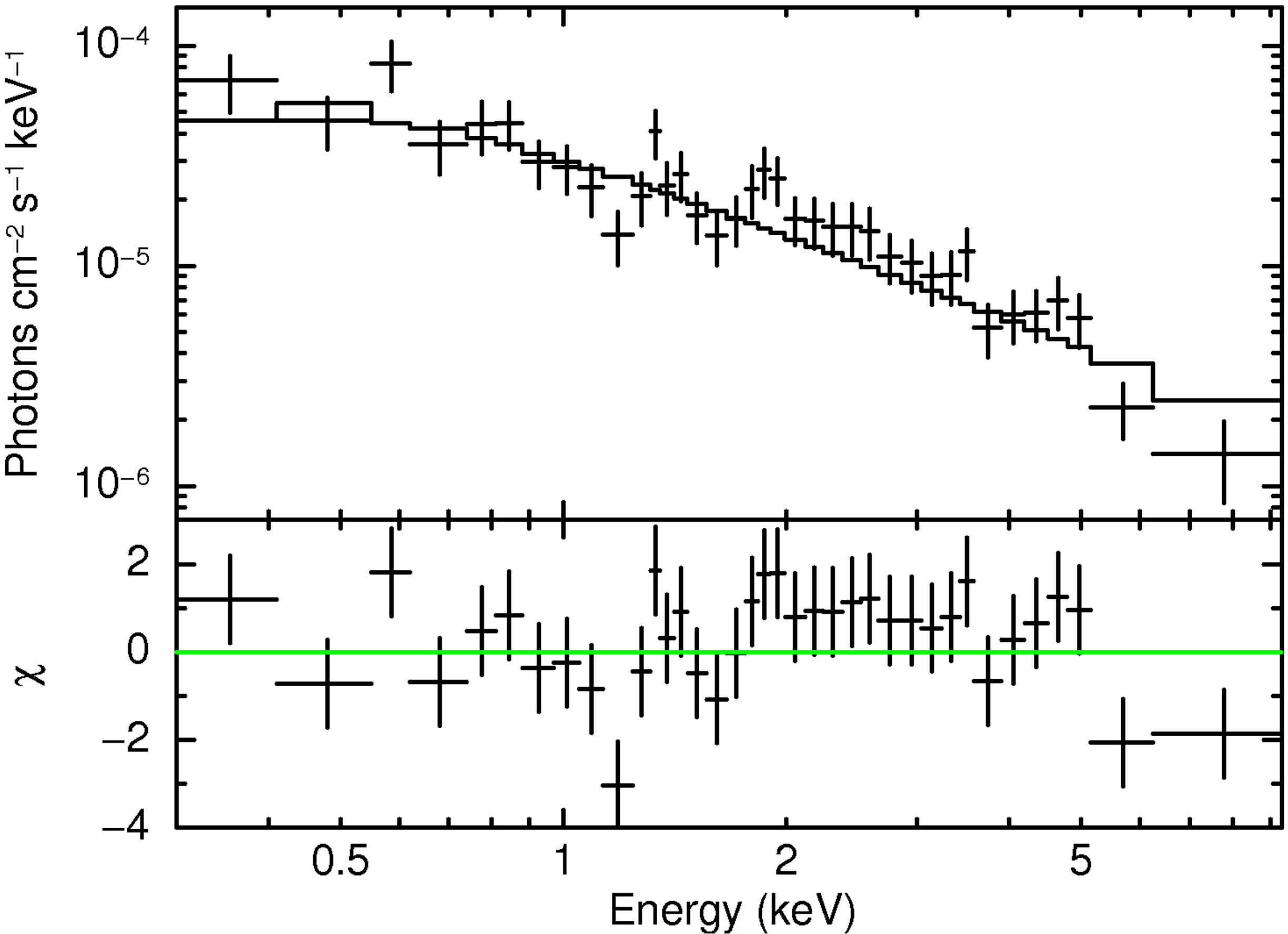}
    \includegraphics[scale=0.31]{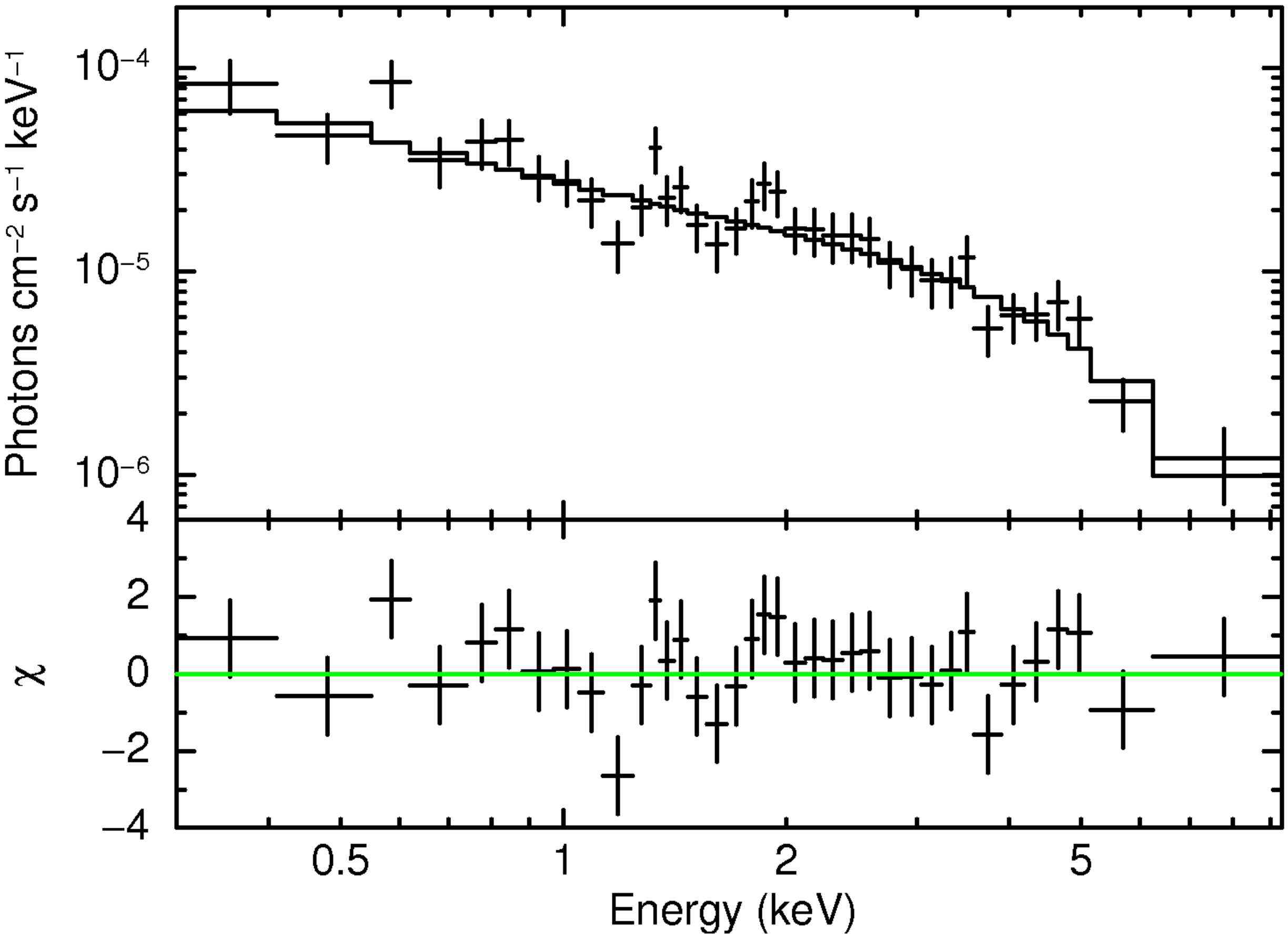}
  \protect\caption{Fit to a power-law (top) and a thermal Comptonization model (bottom) to the \textit{Swift} XRT combined spectrum of N4088--X1. \label{figSwift}}
\end{figure}

We fitted the co-added \textit{Swift} spectrum using {\footnotesize XSPEC} again, with an absorbed power-law model with absorption accounted for using the \textit{wabs} model and the $N_\mathrm{H}$ column set to be greater than the Galactic absorption in the direction of N4088--X1 (i.e. 2 $\times$ 10$^{20}$ atoms cm$^{-2}$). 
The fit obtained, with a hard photon index of 1.3 (see Table~\ref{specfit}), does not satisfactorily describe the observed spectrum, as two data points above 5 keV fall below the power-law fit (see Fig.~\ref{figSwift}, top). We thus attempted to fit this spectrum with \textit{wabs*diskbb} as with our \emph{Chandra} data and an exponential cut-off \textit{wabs*pow*highe}. 
Using the \textit{wabs*diskbb} model, a fit with $\chi^{2}$/dof=39.97/34 and an inner disk temperature of 2.1 keV (see Table \ref{specfit}) is obtained. This temperature is again too high for optically thick emission from a typical optically thin accretion disk. A good fit ($\chi^{2}$/dof=36.89/32) is also obtained with the exponential cut-off model. The high-energy cut-off has been also observed in other ULXs (e.g., \citealt{2006ApJ...641L.125D}, \citeyear{2010MNRAS.407..291D}; \citealt{2006MNRAS.368..397S}; \citealt{2007Ap&SS.311..203R}; \citealt{2009MNRAS.397.1836G}; \citealt{2013MNRAS.435.1758S}) and is well described by thermal Comptonization of hot coronal electrons by soft photons (e.g., \citealt{1994ApJ...434..570T}; \citealt{1995ApJ...449..188H}). We thus test if a Comptonization model (\textit{wabs*comptt}) is able to fit the data (Fig.~\ref{figSwift}, bottom). We find that the \textit{Swift} spectrum is well described ($\chi^{2}$/dof=35.01/32) with a Comptonization model of input soft photon temperature $T_\mathrm{0}<$ 0.1 keV and plasma temperature $kT_\mathrm{e}$=1.2 keV (see Table~\ref{specfit}). The S/N is insufficient for us to detect any contribution from the disk, although the disk should still be present.

Both the \textit{diskbb}, \textit{pow*highe}, and \textit{comptt} models provide significant improved statistics with respect to the \textit{pow} model, as indicated by the $\chi^{2}$/dof. The Bayesian Information Criterion (BIC; \citealt{schwarzBIC}; \citealt{Kass1995}) can be also used as a further indicator of the statistical improvement of one model over another. The BIC value can be calculated as BIC = 2log($L_\mathrm{1}$) - 2log($L_\mathrm{2}$) - ($k_\mathrm{1} - k_\mathrm{2}$)log(n), where $L$ = exp$(-\chi^{2}$/2) for models 1 and 2, respectively, $k$ = number of parameters in model, and $n$ = number of data points. We obtain BIC values in the range 2--6 for the three models compared to the \textit{pow} one, which is a `positive' result\footnote{The significance of a model over another is `preferred' for BIC values between 0--2, `positive' for BIC values between 2--6, `strong' for values 6--10, and `very' strong for BIC $>$ 10}. When comparing the \textit{comptt} to the \textit{diskbb} and \textit{pow*highe} models we obtain BIC numbers between 0 and 2, which indicate that the \textit{comptt} is `preferred' over these models.

To quantify the significance of the spectral cut-off, we also try to fit a broken power-law with the two slopes tied together and then thaw one of them and use the F-test to compare the fits. The F-test assesses whether the improvement of the $\chi^{2}$ is due to chance or to the new component being significant (e.g., \citealt{1989sgtu.book.....B}). 
We obtain a break energy of 4.7$^{+0.6}_{-0.9}$ keV, an F-statistic value of 6.04, and a probability of 0.006, which is $<<$1 and thus indicates that the cut-off is significantly there.


\begin{table*}
\scriptsize
\caption{Spectral parameters obtained by fitting the \emph{Swift} XRT combined spectrum of N4088--X1 with an absorbed power-law, power-law with exponential cut-off, disk-blackbody, and thermal Comptonization model. The quoted flux is the observed flux in the 0.2 to 10.0 keV energy range. The errors provided are at 90\% confidence level. \label{specfit}} 
\begin{tabular}{lccccccccc}
\tableline
Model  & $N_\mathrm{H}$ & $\Gamma/T_\mathrm{in}$ & Cut-off   &   $T_\mathrm{0}$   &     $kT_\mathrm{e}$    &  $\tau$ &  Norm. & Flux & $\chi^{2}$/dof \\
 & (10$^{22}$ cm$^{-2}$) & ( /keV) & (keV) &   (keV)  &   (keV) &  & &(erg cm$^{-2}$ s$^{-1}$) &\\
\tableline
\textit{pow} & 0.05$^{+0.06}_{-0.05}$ & 1.3$^{+0.2}_{-0.2}$ &  --   &  -- &   -- & -- &(3.4$^{+0.8}_{-0.6}$) $\times$ 10$^{-5}$ & 3.7$^{+0.4}_{-0.3}$ $\times$ 10$^{-13}$ & 51.06/34 \\
\textit{pow*highe} & 0.02$^{+0.03}_{-0.02}$ & 0.9$^{+0.2}_{-0.3}$ & 3$^{+2}_{-1}$  & -- & -- & -- & (2.9$^{+0.4}_{-0.3}$) $\times$ 10$^{-5}$ & (3.0$^{+0.4}_{-0.4}$) $\times$ 10$^{-13}$ & 36.89/32 \\
\textit{diskbb} & 0.02$^{+0.02}_{-0.02}$ & 2.1$^{+0.3}_{-0.4}$ &  --   & -- & -- & --& (8$^{+7}_{-3}$) $\times$ 10$^{-4}$ & (3.0$^{+0.4}_{-0.3}$) $\times$ 10$^{-13}$ & 39.97/34 \\
\textit{comptt} & 0.02$^{+0.03}_{-0.02}$ & 	--		&	--	&  0.03$^{+0.1}_{-0.03}$  &   1.2$^{+0.2}_{-0.2}$   & 17$^{+3}_{-3}$ &   (1.3$^{+1.0}_{-0.5}$) $\times$ 10$^{-4}$ &  (2.9$^{+0.3}_{-0.3}$) $\times$ 10$^{-13}$ & 35.01/32 \\ 
\tableline
\end{tabular}
\end{table*}

\subsection{Radio counterpart of the ULX}
\label{radiocounterpart}
A 1.6 GHz radio counterpart to N4088--X1 consistent with the \textit{ROSAT} X-ray detection was reported by \cite{2011AN....332..379M}. The \textit{Chandra} position of N4088--X1 obtained here reveals an offset of 6 arcsec between the X-ray position and the 1.6 GHz EVN radio source of \cite{2011AN....332..379M}, which can now be ruled out as a candidate counterpart of the ULX.

\begin{figure}[h!]
 \hspace{-18pt}
\includegraphics[scale=0.49]{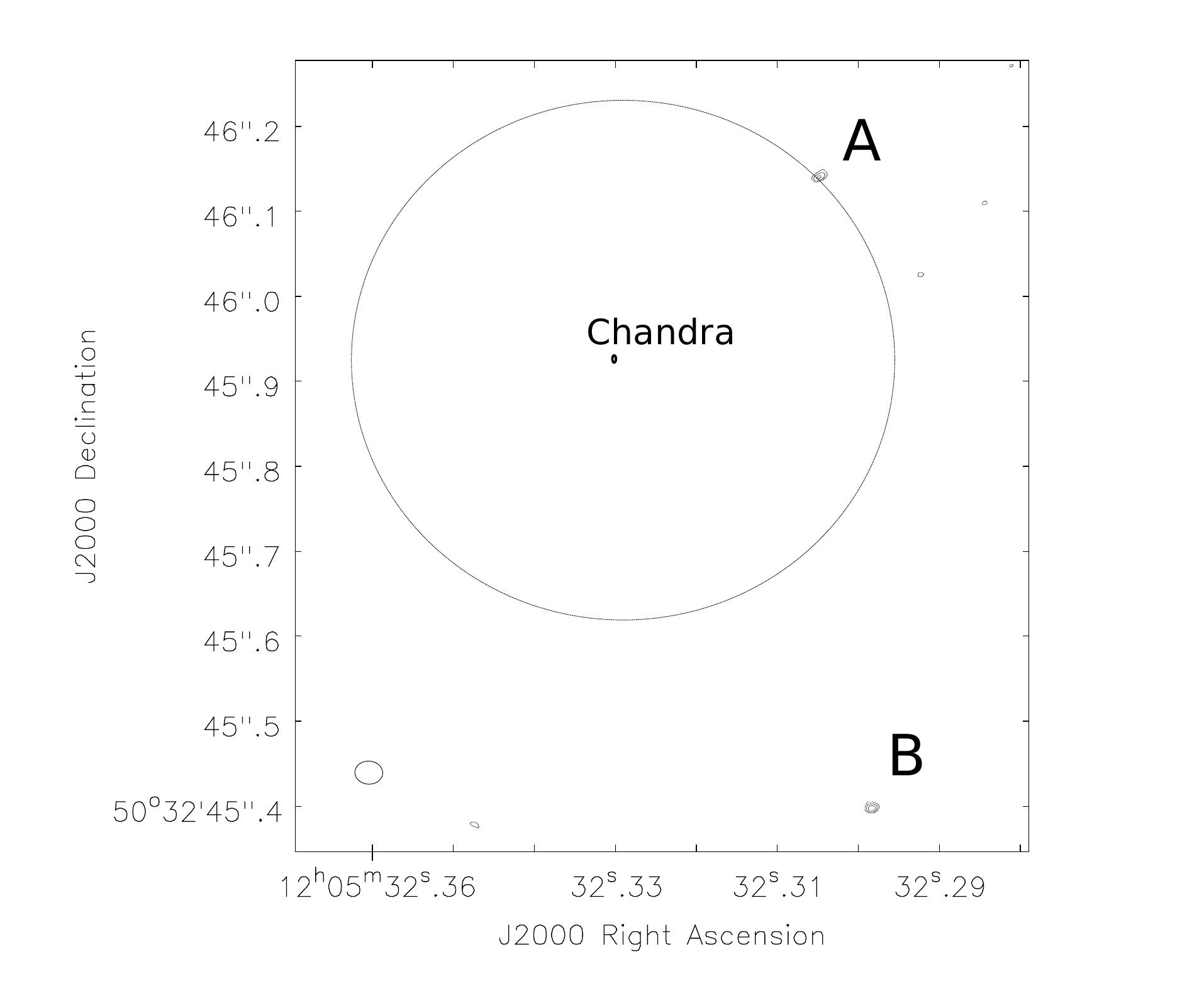}
  \protect\caption{1.6 GHz EVN image at the \textit{Chandra} position of N4088--X1. The contours are (3, 4, 5) times the off-source rms noise of 11 $\mu$Jy beam$^{-1}$. Two compact components labelled A and B are detected. A circle of  radius 0.3 arcsec centered at the \textit{Chandra} position is plotted to ease the visualization of the distance between the ULX and the radio components. The beam size shown at the bottom left corner is 33 mas $\times$ 27 mas oriented at a position angle of 86$^{\circ}$.4.\label{radio}}
\end{figure}

The reanalysis of the 1.6 GHz using a FOV of 1 arcsec around the \textit{Chandra} position yields the detection of compact radio emission consistent, within 0.3 arcsec, with the \textit{Chandra} positional error (component A in Fig.~\ref{radio}). This compact component A is detected at a 5.2$\sigma$ level and is centered at RA(J2000) = 12$^h$05$^m$32$^s$.3048 $\pm$ 0.0004$^s$, Dec.(J2000) = 50$^{\circ}$32\arcmin46\arcsec.140 $\pm$ 0.004\arcsec. It has an integrated flux density of 49 $\mu$Jy, from which we derive a 1.6 GHz luminosity $L_\mathrm{1.6 GHz}= 1.6 \times 10^{34}$ erg s$^{-1}$.
The AIPS task JMFIT is used to fit an elliptical Gaussian to this peak of emission, which yields a lower limit on the brightness temperature T$_\mathrm{B}$ $>$ 3 $\times$ 10$^{4}$ K.

A second component (labeled B in Fig.~\ref{radio}) is detected at 5.4$\sigma$ level offset 0.6 arcsec from the \textit{Chandra} X-ray position. Component B is centered at RA(J2000) = 12$^h$05$^m$32$^s$.2983 $\pm$ 0.0003$^s$, Dec.(J2000) = 50$^{\circ}$32\arcmin45\arcsec.398 $\pm$ 0.003\arcsec, and has an integrated flux density of 55 $\mu$Jy, from which we derive a 1.6 GHz luminosity $L_\mathrm{1.6 GHz}= 1.8 \times 10^{34}$ erg s$^{-1}$ assuming it is in NGC\,4088. The fit of an elliptical Gaussian to this peak of emission yields a lower limit on the brightness temperature T$_\mathrm{B}$ $>$ 3 $\times$ 10$^{4}$ K.

In the 5 GHz EVN observations, no radio emission is detected above a 5$\sigma$ level for N4088--X1 within 1 arcsec of the \textit{Chandra} position nor within a FOV of 5 arcsec around the VLA position. An upper limit on the flux density of the ULX of 0.30 mJy beam$^{-1}$ is obtained by estimating the rms at the \textit{Chandra} position, from which we derive an upper limit on the brightness temperature $T_\mathrm{B} < 6 \times 10^{5}$ K. Adopting a distance to NGC\,4088 of 13 Mpc, we derive an upper limit on the 5 GHz radio luminosity of $L_\mathrm{5 GHz} < 3.1 \times 10^{35}$ erg s$^{-1}$.

Combined with the 1.6 GHz detection, the upper limit on the flux density at 5 GHz can be used to constrain the spectral index of the source. Defining $F_{\nu}\propto\nu^{\alpha}$, we obtain $\alpha\leq1.6$, which is trivially satisfied by any physical class of radio spectra. It should be noted that this spectral index is derived from non-simultaneous observations and can thus be affected by variability effects (unless changes in the radio emission occur on timescales longer than years). Therefore, deeper 5 GHz observations simultaneous with new 1.6 GHz ones are needed before we can determine the nature of the radio emission.

\subsection{X-ray luminosity function}
\subsubsection{AGN fraction}
In order to estimate the level of AGN contamination, we compared the flux distribution of the \textit{Chandra} sources with the expected AGN flux distribution (\citealt{2012ApJ...752...46L}). We estimate that about 2/3 of the thirty-one \textit{Chandra} sources detected inside the full S3 chip (area = 0.0196 deg$^{2}$) are likely to be AGN. However, for the fifteen sources inside the D25 ellipse (area = 0.0028 deg$^{2}$), the AGN fraction is only $\sim$19\% (i.e. $\leq$ 3 AGN). In particular, the probability of finding an AGN inside the D25 with the flux measured for the ULX is $\sim$15\%. Although the probability of detecting a background AGN inside the D25 of NGC\,4088 is not very low, other considerations argue against this possibility for the ULX (see Section~\ref{discussion}).

\subsubsection{XLF fitting}
\label{XLF}
\begin{figure}
 \hspace{-17pt}
 \includegraphics[scale=0.5]{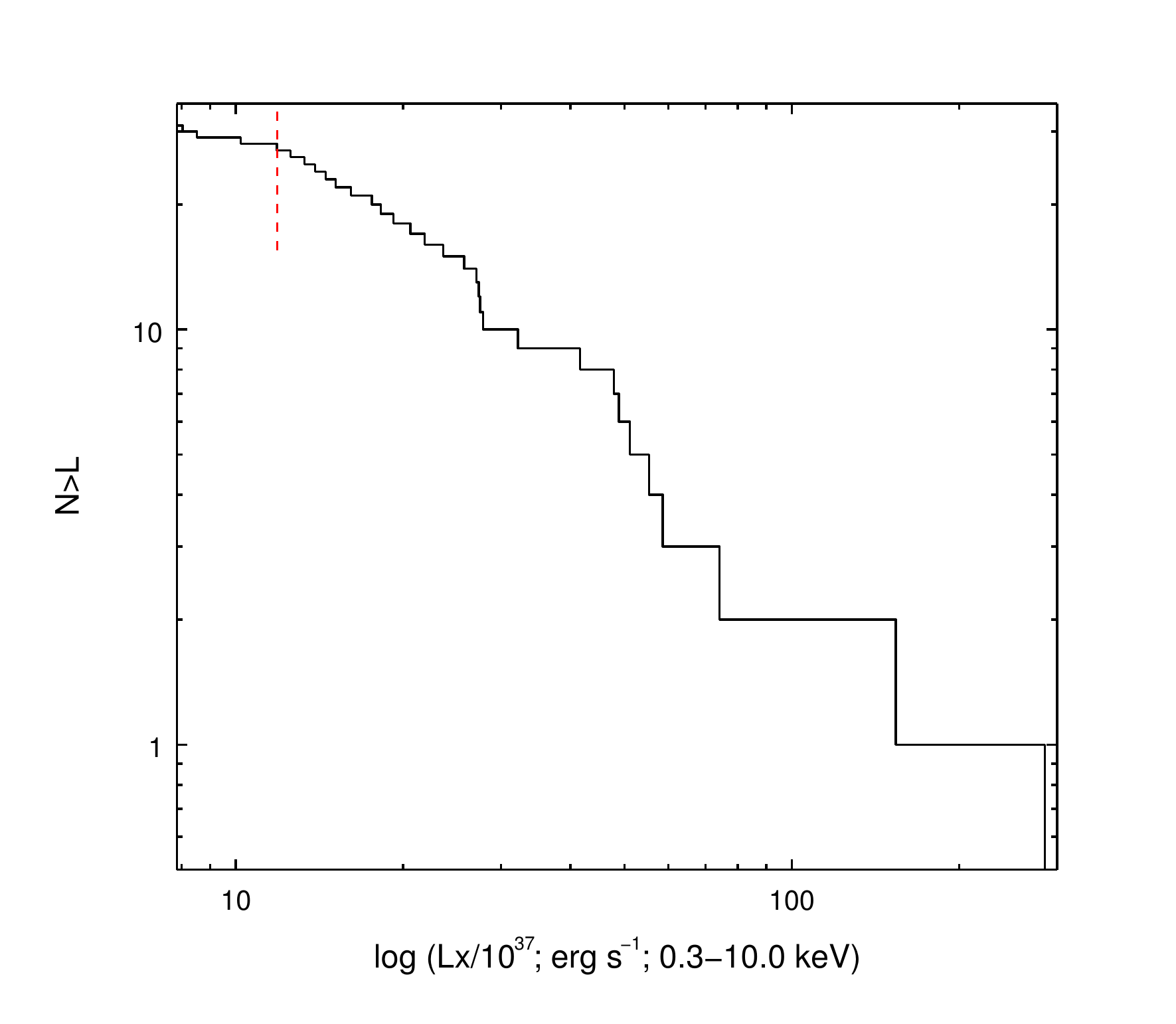} 
  \protect\caption{Cumulative XLF of NGC\,4088. The vertical dotted red line shows the cut performed in the XLF to correct for incompleteness.\label{cumXLF}}
\end{figure}

The XLF of the X-ray source population of NGC\,4088 is constructed considering those sources inside the D25 ellipse of the galaxy.
An apparent flattening of the XLF is observed at the low luminosities (see Fig.~\ref{cumXLF}), which may be caused by incompleteness effects. To correct for this, we eliminate the three lower points of the XLF that may be affected by incompleteness (it should be noted that the purpose of this paper is not to study the low-luminosity XLF of NGC\,4088). As a result, the XLF flattening disappears.

Since the errors in the cumulative XLF are not independent, from now on we use the differential XLF. This allows us to take into account the statistical uncertainties, which are large given the very small number of sources. These errors do not take into account the possibility that $\leq$ 3 sources inside the D25 are AGN.

\begin{figure}
 \hspace{-20pt}
 \includegraphics[scale=0.45]{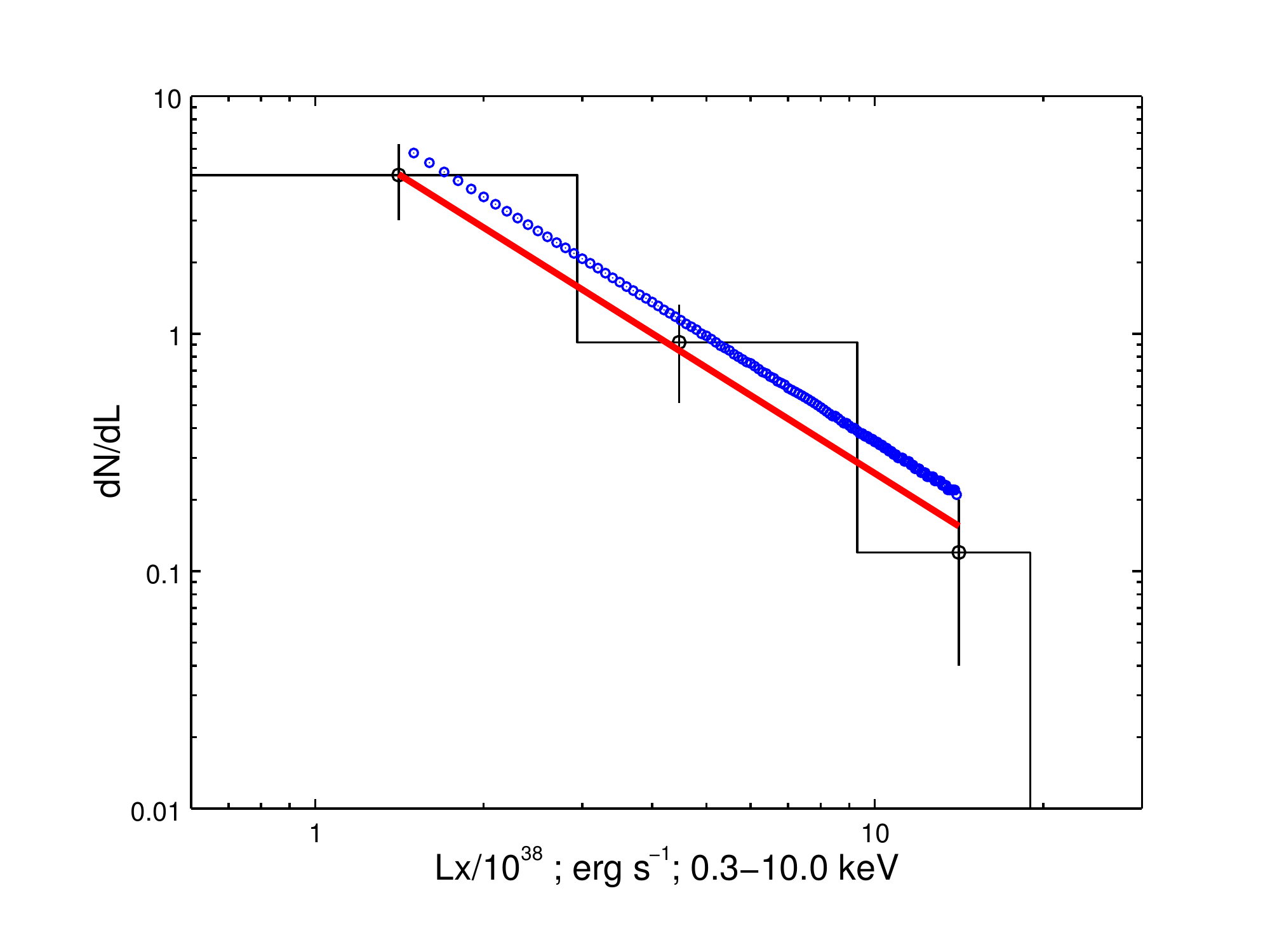}
  \protect\caption{Fit of the differential XLF inside the D25 ellipse of NGC\,4088 to a single power-law (red solid line). The best-fit slope is $\alpha=0.5 \pm 0.8$. The blue dotted line denotes the XLF of star-forming galaxies from \cite{2012MNRAS.419.2095M} scaled by a factor 4.5. \label{diffXLF}} 
\end{figure}

We fit the differential XLF with a power-law of the form $dN/dL_\mathrm{X}=BL_\mathrm{X}^{-\beta}$ normalized to $10^{37}$ erg s$^{-1}$, where $\beta=\alpha+1$ (Fig.~\ref{diffXLF}). The Cash statistic (\citealt{1979ApJ...228..939C}) is used instead of the minimum $\chi^{2}$ method due to insufficient data points for the errors to be described by Gaussian statistics. 
The fit gives a slope of $\beta=1.5\pm 0.8$ (1$\sigma$ error), and thus $\alpha=0.5 \pm 0.8$, which is similar to the slope found in other spiral galaxies (e.g., $\alpha\sim0.4-0.5$; the Antennae galaxies, \citealt{2002ApJ...577..726Z}; M82, \citealt{2007ApJ...661..135Z}; M81, \citealt{2001ApJ...549L..43T}). In these galaxies, the X-ray source population is mostly dominated by HMXBs. The ULX N4088--X1, with a luminosity $L_\mathrm{0.2-10.0 keV} = 3.4\ \times$ 10$^{39}$ erg s$^{-1}$, could thus be located at the high end of the HMXB distribution of NGC\,4088, which is consistent with the association of ULXs with young stellar populations (e.g., \citealt{2002ApJ...577..726Z}; \citealt{2004ApJS..154..519S}). However, the possibility that the ULX is a LMXB (e.g., \citealt{2012MNRAS.420.2969M}; \citealt{2012ApJ...750..152S}) cannot be ruled out.

To further study the star-formation rate (SFR) of NGC\,4088, we overplot the XLF of star-forming galaxies from \cite{2012MNRAS.419.2095M}, fig.~4, black line, on the XLF of NGC\,4088. 
These authors normalize their XLF to 1 $M_{\odot}$ yr$^{-1}$. By rescaling this value to fit our data (see Fig.~\ref{diffXLF}), it is possible to obtain an estimate on the SFR for NGC 4088. This provides an estimated rate of 4.5 $M_{\odot}$ yr$^{-1}$, in good agreement with the range of SFRs 1.7--7.8 $M_{\odot}$ yr$^{-1}$ obtained by \cite{2006ApJ...643..173S} in the H$\alpha$, IR, and radio bands.

\subsection{The supernova SN2009dd}
Based on \textit{Swift} XRT observations, \cite{2013A&A...555A.142I} reported the X-ray detection of a recent type II supernova (SN2009dd) in the galaxy NGC\,4088 at RA(J2000) = 12$^h$05$^m$34$^s$.10, Dec.(J2000) = 50$^{\circ}$32\arcmin19\arcsec.4. SN2009dd brightened from 8 $\times$ 10$^{38}$ erg s$^{-1}$ to 1.7 $\times$ 10$^{39}$ erg s$^{-1}$ in the 0.2--10 keV energy range.
While \textit{XMM-Newton} does not have enough resolution to resolve the SN from the nucleus, the \textit{Chandra} observations of NGC\,4088 here presented can provide the most accurate X-ray position of SN2009dd. 

In our \textit{Chandra} observations, \textit{wavdetect} fails to detect the SN2009dd. However, the source can be dimly seen in the 0.3--10 keV and 1.5--7.0 keV bands. Using CIAO Statistics, we obtain $\sim$8 counts in the 0.3--10 keV background subtracted image, which corresponds to an unabsorbed flux $F_\mathrm{0.3-10 keV}$ = 4.2 $\times$ 10$^{-15}$ erg cm$^{-2}$ s$^{-1}$ and luminosity $L_\mathrm{0.3-10 keV}$ = 8.4 $\times$ 10$^{37}$ erg s$^{-1}$ assuming $\Gamma = 1.8$ and Galactic column density $N_\mathrm{H} = 2 \times 10^{20}$ cm$^{-2}$. From the detected number of source counts and background counts in the source extraction area, we can estimate the 90\% confidence limit to the true number of counts coming from SN2009dd from Poisson statistics (e.g., \citealt{1991ApJ...374..344K}). We obtain a 90\%--confidence level lower limit $\sim$3 and upper limit $\sim$13, which indicates that the source is real and not a background fluctuation. The chance probability of detecting $\geq$ 8 counts for the given background is $\sim10^{-5}$.
The \textit{Chandra} location of SN2009dd is RA(J2000) = 12$^h$05$^m$34$^s$.08, Dec.(J2000) = 50$^{\circ}$32\arcmin19\arcsec.0, which is consistent with the \textit{Swift} position within the $\sim$0.6 arcsec \textit{Chandra} absolute astrometry.

No radio emission was detected at the position of SN2009dd above a 3$\sigma$ upper limit of 0.35 mJy at 1.3 cm and of 0.15 mJy at 3.5 cm according to \cite{2009ATel.2016....1S}, and there are no other recent observations of NGC\,4088 apart from the ones reported in 2009.
The VLBI observations presented in this paper are centered too far from the position of SN2009dd; hence, no further upper limits to the radio flux density can be provided.

\section{Discussion}
\label{discussion}
\subsection*{The nature of N4088--X1}
The results of the XLF fitting (Section~\ref{XLF}) show that the XLF of NGC\,4088 is well described by a power-law of $\alpha$=0.5, which resembles the typical fits of HMXB XLFs and indicates that N4088--X1 could be located at the high-luminosity extension of the BH XRB distribution. 
These results are in agreement with studies of the XLF of star-forming galaxies (e.g., \citealt{2003MNRAS.339..793G}; \citealt{2012MNRAS.419.2095M}) and of the location of ULXs in stellar clusters (e.g., \citealt{2013MNRAS.432..506P}), which conclude that ULXs are a high-luminosity end of the XRB population harboring most possibly stellar-mass BHs rather than IMBHs.

When analyzing the \textit{Chandra} X-ray spectrum of N4088--X1 (Section~\ref{xrayprop}), we find that it is acceptably fitted both by a simple absorbed power-law continuum of $\Gamma$=1.1$^{+0.6}_{-0.5}$
and by a disk-blackbody model with $kT_\mathrm{in}$=2.5 keV. The hard power-law slope is consistent with the classification of N4088--X1 as a ``hard ULX" (\citealt{2011AN....332..330S}; \citealt{2013MNRAS.435.1758S}). The physical interpretation of hard ULXs is still unclear: some authors (e.g., Winter et al. 2006) suggested that they are IMBHs in the low/hard state; others (e.g., \citealt{2009MNRAS.397.1836G}; \citealt{2011AN....332..330S}; \citealt{2013MNRAS.435.1758S}) explain them instead as one possible variety of super-Eddington accretion (e.g., \citealt{2007MNRAS.377.1187P}; \citealt{2009MNRAS.393L..41K}). The photon index is much harder than the value of $\Gamma \sim$1.7 that is typical of the low/hard state (e.g., Remillard \& McClintock 2006) but consistent with it within the 90\% margin of error, which does thus not rule out the interpretation of N4088--X1 as an IMBH. A disk temperature $>$ 1 keV is however not consistent with an IMBH but favors the nature of N4088--X1 as an XRB with super-Eddington accretion. The low number of detected counts are insufficient to statistically distinguish between the models. 

In the spectral fitting performed on the \textit{Swift} data, a disk-blackbody model with $kT_\mathrm{in}\sim$2 keV provided a better fit than the power-law model. A hint of a spectral cut-off at $\sim$5 keV seems to be observed in the residuals of the power-law model. This feature is similar to that observed in other ULXs (e.g., \citealt{2006ApJ...641L.125D}, 2010; \citealt{2006MNRAS.368..397S}; \citealt{2012MNRAS.422..990K}; \citealt{2013MNRAS.435.1758S}), suggesting that N4088--X1 could be in the ultraluminous state (e.g., \citealt{2007Ap&SS.311..203R}; \citealt{2009MNRAS.397.1836G}; \citealt{2009MNRAS.398.1450K}; \citealt{2013MNRAS.435.1758S}) with a thermal spectrum described by Compton scattering of soft photons (e.g., \citealt{2012MNRAS.420.2969M}; \citealt{2012ApJ...750..152S}; \citealt{2013A&A...553A..61S}). The good fit provided by the thermal Comptonization model together with the statistically significant cut-off indicated by the F-test support this possibility, although the \textit{Swift} data are not of high enough S/N to confirm it. Therefore, whilst the spectrum is merely quasi-thermal in shape, we are unable to confirm the accretion state at this time using the X-ray data at hand. However, we can place constraints on the nature of N4088--X1 when combining these fits with information from other wavebands.

In Section~\ref{radiocounterpart} we have reported the 1.6 GHz EVN detection of two compact components within the \textit{Chandra} error circle of 0.6 arcsec. These could be either compact HII regions (e.g., see review by \citealt{2002ARA&A..40...27C}), compact SNRs, or an accreting BH. Compact HII regions have typical sizes of 1--7 pc, a thermal X-ray spectrum with temperatures $>$ 2 keV, an inverted radio spectrum of spectral index $\sim$1, and $T_\mathrm{B} < 10^{4}$ K (e.g., \citealt{2001ApJ...559..864J}; \citealt{2002MNRAS.334..912M}; \citealt{2006ApJ...653..409T}; \citealt{2007prpl.conf..181H}). Compact SNRs have also typical sizes of a few pc and a steep radio spectrum (e.g., \citealt{2007AJ....133.2156L}; \citealt{2013MNRAS.436.2454M}b). 
The limits on the brightness temperatures derived from the 1.6 GHz detections and the 5 GHz non-detection ($3 \times 10^{4} < T_\mathrm{B} < 6 \times 10^{5}$ K) together with the high X-ray luminosity of N4088--X1 make the presence of compact HII regions quite improbable and are more indicative of the presence of either compact SNRs or an accreting BH. The EVN beam size of $\sim$30 mas (which corresponds to $\sim$2 pc at the distance of the galaxy) is consistent with both a compact HII region, a compact SNR, and an accreting BH.
Unfortunately, the non-detection at 5 GHz and the upper limit on the radio spectral index ($\alpha\leq$1.6) estimated for N4088--X1 does not clarify whether the radio counterpart of this ULX is steep, flat, inverted, or variable. We plan to obtain deeper radio observations to determine the spectral index and variability properties of this source and therefore constrain the physical interpretation. 

In order to test if the two radio components are associated with the ULX, we derive the probability of a chance alignment between the \textit{Chandra} counterpart and any compact radio source in the 5 arcsec field (positional error of the VLA) in that region of the galaxy. We use the number of sources detected above 5$\sigma$ in the imaged VLA field (see Section~\ref{VLBI}) and the \textit{Chandra} error circle of 0.6 arcsec following the same approach as in \citeauthor{2013MNRAS.436.1546M} (2013a). This gives a probability of chance alignment P(CA)=0.4, which is quite high and indicates that one or the two detected compact radio components (A and B, Fig.~\ref{radio}) could correspond to a random source (i.e. a compact SNR or BH) not associated with the ULX. Given the 0.6 arcsec offset between component B and the \textit{Chandra} position, component B is most plausibly not related to N4088--X1 while the radio emission of component A (consistent with the \textit{Chandra} position within 0.3 arcsec) could be coming from a BH associated with the ULX.
In this case, the emission of component A could be due to flaring radio emission from a ballistic jet (e.g., \citealt{2012Sci...337..554W}; \citealt{2013Natur.493..187M}) or compact core emission if the source is in a low/hard state (albeit, as mentioned above, with a flatter photon spectral index than typical).
The results of the X-ray spectral analysis suggest that N4088--X1 is in a thermal (i.e. Comptonized) ultraluminous state. Therefore, if the radio emission is associated with the ULX then the compact radio component is most likely associated with ballistic jet emission. 

If we assume for a moment that the source is residing in the low/hard state accreting at $L <$ 10\% Eddington (e.g., \citealt{2003MNRAS.342.1041D}), we are able to invoke the fundamental plane of accreting BHs (e.g., \citealt{2003MNRAS.345.1057M}; \citealt{2006A&A...456..439K}; \citealt{2012MNRAS.423..590G}) and estimate a BH mass. 
For this, we use the 2--10 keV X-ray flux obtained from the \textit{Chandra} spectral fitting and scale the 1.6 GHz flux density to 5 GHz using a radio spectral index $\alpha = 0.15$ (a typical spectral index for flat cores used to estimate the BH mass from the Fundamental Plane, e.g., \citealt{2004A&A...414..895F}).
Using the Fundamental Plane of \cite{2006A&A...456..439K} that presents the least scatter, an upper limit on the BH mass of $3 \times 10^{5} M_{\odot}$ is obtained. This is consistent with this source being either an IMBH or an XRB, and rules out the nature of N4088--X1 as a SMBH.

We also derive the ratio $R_\mathrm{X}$ of 5 GHz radio emission to 2--10 keV X-ray emission defined by \cite{2003ApJ...583..145T}. Typical values of this ratio for XRBs are $R_\mathrm{X} < -5.3$ (e.g., \citealt{2011Natur.470...66R}), while values of $R_\mathrm{X}=-2.7$ to $-2$ have been estimated in SNRs (e.g., \citealt{2003ApJ...599.1043N}; \citealt{2013MNRAS.436.2454M}b). For low-luminosity AGN (LLAGN), Ho (2008) reports a range of values $-3.8 < R_\mathrm{X} < -2.8$ (see also \citealt{2013MNRAS.436.1546M}a, table~3). For N4088--X1, we obtain $R_\mathrm{X} < -4.7$ using the 2--10 keV X-ray flux obtained from the \textit{Chandra} spectral fitting and the 5 GHz- scaled radio luminosity. This is in agreement with the previous results, ruling out both a SMBH and a SNR nature for this ULX.

The location of N4088--X1 in the spiral arm of the host galaxy, possibly within an HII region (e.g., \citealt{2006A&A...452..739S}), and the low X-ray absorption seen in the \textit{Chandra} and \textit{Swift} data also argue very strongly against a LLAGN background scenario. On the other hand, the lack of a bright counterpart to N4088--X1 in the optical image makes it very unlikely to be a foreground star.

\section{Conclusions}
\label{conclusions}
We have presented the first \textit{Chandra} and \textit{Swift} X-ray observations of the galaxy NGC\,4088 and the ULX N4088--X1 that it hosts. 
EVN observations at 1.6 and 5 GHz of the ULX were performed almost simultaneously to the \textit{Swift} and \textit{Chandra} observations, respectively, which have allowed us to investigate the compact radio emission of a ULX radio counterpart previously proposed with the VLA.

The X-ray spectral analysis of N4088--X1 seems to favor a thermally Comptonized spectrum for this source, although the possibility that it is a hard ULX cannot be ruled out. The disk temperature ($kT_\mathrm{in}\sim$2 keV) obtained from the disk-blackbody model and the presence of a statistically significant spectral break at $\sim$5 keV are not consistent with N4088--X1 being an IMBH  but suggest that the source could be an XRB in a super-Eddington ultraluminous state. Unfortunately, this cannot be confirmed with the present data due to the low S/N. 
If the source is in an ultraluminous state, the detection of compact radio emission at 1.6 GHz coincident with the \textit{Chandra} counterpart could then correspond to ballistic jet emission from an accreting BH. Multi-epoch multi-wavelength observations are required to confirm this.

Finally, the detection of fifteen sources within the D25 ellipse of NGC\,4088 has allowed us to fit the XLF of this galaxy and estimate a SFR of 4.5 $M_{\odot}$ yr$^{-1}$. We find that the XLF resembles that of typical star-forming galaxies, where the ULX N4088--X1 could be at the high-luminosity end of the XRB population. We thus conclude that N4088--X1 is possibly a HMXB with a thermally Comptonized spectrum and either approaching the Eddington limit or in the ultraluminous state.

\section{Acknowledgements}
This work was partially supported by the Chandra X-Ray Center (CXC), which
is operated by the Smithsonian Astrophysical Observatory (SAO) under NASA contract NAS8-03060, and by Chandra Director Discretionary Time grant DD2-13063X. MM acknowledges finantial support by PAYA2o11-25527

J.C.G. would like to acknowledge Avadh Bhatia Fellowship, the Alberta Ingenuity New Faculty Award, and the financial support from NSERC Discovery Grants. 
SAF is the recipient of an Australian Research Council Postdoctoral Fellowship, funded by grant DP110102889.
RS acknowledges support from the Australian Research CouncilÕs Discovery Projects funding scheme (project number
DP120102393).

\bibliographystyle{mn2e} 
\bibliography{referencesALL}

\end{document}